\documentclass[pra,aps,amsmath,amssymb,showkeys]{revtex4}
\usepackage{graphicx}
\newcommand{\hh}{{\mathcal{H}}}

\newcommand{\pen}{\openone}

\newcommand{\tr}{\mathrm{tr}}

\newcommand{\bro}{\boldsymbol{\rho}}

\newcommand{\cmb}{\mathbb{B}}
\newcommand{\clb}{{\mathcal{B}}}
\newcommand{\cle}{{\mathcal{E}}}
\newcommand{\clf}{{\mathcal{F}}}

\newcommand{\cln}{{\mathcal{N}}}
\newcommand{\mf}{\mathsf{F}}

\newcommand{\cleg}{{\mathcal{G}}}
\newcommand{\hlm}{{\mathsf{G}}}

\newcommand{\nm}{{\mathsf{N}}}
\newcommand{\me}{{\mathsf{E}}}

\newcommand{\xdif}{{\mathrm{d}}}
\newcommand{\mgb}{\mathbb{E}}
\newcommand{\mnn}{\mathbb{N}}

\begin{document}
\clearpage
\preprint{}

\title{Uncertainty relations in terms of generalized entropies derived from information diagrams}

\author{Alexey E. Rastegin}
\affiliation{Department of Theoretical Physics, Irkutsk State University, K. Marx St. 1, Irkutsk 664003, Russia}

\begin{abstract}
Entropic uncertainty relations are interesting in their own rights
as well as for a lot of applications. Keeping this in mind, we try
to make the corresponding inequalities as tight as possible. The use
of parametrized entropies also allows one to improve relations
between various information measures. Measurements of special types
are widely used in quantum information science. For many of them we
can estimate the index of coincidence defined as the total sum of
squared probabilities. Inequalities between entropies and the index
of coincidence form a long-standing direction of researches in
classical information theory. The so-called information diagrams
provide a powerful tool to obtain inequalities of interest. In the
literature, results of such a kind mainly deal with standard
information functions linked to the Shannon entropy. At the same
time, generalized information functions have found use in questions
of quantum information theory. In effect, R\'{e}nyi and Tsallis
entropies and related functions are of a separate interest. This
paper is devoted to entropic uncertainty relations derived
from information diagrams. The obtained inequalities are then
applied to mutually unbiased bases, symmetric informationally
complete measurements and their generalizations. We also improve
entropic uncertainty relations for quantum measurement assigned to
an equiangular tight frame.
\end{abstract}


\maketitle

\pagenumbering{arabic}
\setcounter{page}{1}

\section{Introduction}\label{sec1}

Heisenberg's uncertainty principle \cite{heisenberg} emphasizes a
fundamental scientific concept, though originally it was posed in
quantum theory. In fact, the observer changes any system that he
tries to explore. Uncertainty relations are interesting not only as
related to the cornerstones of quantum mechanics. Such relations
typically show feasible schemes to detect non-classical correlations
with an impact on questions of quantum information. It is
insufficient to deal here with the traditional way to pose
uncertainty relations in terms of the product of variances
\cite{kennard,robert}. An information-theoretic point of view is
fruitfully used in various topics including quantum theory
\cite{graaf}. Entropic relations have been proposed as a powerful
tool to characterize quantum uncertainties. For finite-dimensional
observables, basic results in this direction are reviewed in
\cite{ww10,cbtw17}. Entropic uncertainty relations for continuous
variables are considered in \cite{brud11,cerf19}. The entropic
approach allows one to formulate uncertainty relations in the
presence of quantum memory \cite{bccrr10}. Noise-disturbance
formulations \cite{bhow14,rastnos} and relations for the scenario of
successive measurements \cite{bfs14,zzyu15,rastannb} can be posed in
terms of conditional entropies.

There exist special types of measurements are particularly important
in quantum information science. Mutually unbiased bases \cite{bz10}
and symmetric informationally complete measurements \cite{rbksc04}
are used for many purposes. Each symmetric informationally complete
measurement realizes a complex equiangular tight frame with the
maximal number of elements \cite{waldr2018}. These topics are
examples of very difficult questions abouit existence of certain
discrete structures in Hilbert space (see, e.g., the papers
\cite{fhs2017,fivo20} and references therein). For example, the
maximal number of mutually unbiased bases remains unknown even for
six dimensions, i.e., for the smallest dimension that is not a prime
power. The answer is known for mutually unbiased measurements
proposed in \cite{kalev14}. Namely, the complete set of $d+1$ such
measurements in $d$-dimensional Hilbert space always exists
\cite{kalev14}. It is known as Zauner's conjecture \cite{zauner11}
that symmetric informationally complete measurement exist for all
dimensions. When the consideration is not restricted to rank-one
operators, the existence has been proved in \cite{gour2014}.
Information-theoretic uncertainty relations give a tool to detect
non-local correlations such as entanglement and quantum
steerability.

The measurements listed above are such that their
 indices of coincidence can be evaluated or estimated.
To convert the obtained restrictions into entropic inequalities, we
need to describe allowed entropic values in terms of the index of
coincidence. For the Shannon entropy, this question was resolved by
the authors of \cite{harr2001} who extensively used the so-called
information diagrams. The contribution of this paper is three-fold.
First, the lower bound on the Shannon entropy given in
\cite{harr2001} is extended to R\'{e}nyi and Tsallis entropies. It
can be interesting in various questions including applications
mentioned in \cite{harr2001}. Second, information diagrams for the
maximal probability versus the index of coincidence are explicitly
described. Third, we reconsider uncertainty relations in terms of
generalized entropies for quantum measurements listed in the
previous paragraph. The paper is organized as follows. Section
\ref{sec2} reviews the preliminary concepts and fixes the notation.
The reasons to accomplish the investigation are presented in
Section \ref{sec3}. The derivation of the main result is presented
in Section \ref{sec4}. New entropic uncertainty relations for
quantum measurements with special properties are considered in
Section \ref{sec5}. Section \ref{sec6} concludes the paper.

\section{Definitions and notation}\label{sec2}

In this section, the required material is reviewed. First, entropic
functions of interest are discussed. Second, we recall some special
types of quantum measurements. In this paper, probability
distributions are assumed to have a fixed finite support. By
$M_{+}^{1}(n)$, we mean the set of all distributions
$P=(p_{1},\ldots,p_{n})$ on the $n$-set $\{1,\ldots,n\}$. Running
$n\in\mnn$ gives the set of discrete probability distributions
$M_{+}^{1}(\mnn)$. Following \cite{harr2001}, the notation $U_{k}$
is used for the generic uniform distribution over a $k$-set:
\begin{equation}
U_{k}=\biggl(\frac{1}{k}\,,\ldots,\frac{1}{k}\,,0\,,\ldots\biggr)
\, . \label{ukdf}
\end{equation}
To characterize the given probability distribution, we will utilize
the Tsallis \cite{tsallis} and R\'{e}nyi entropies \cite{renyi61}.
These parametrized entropies have found use in various disciplines.
For $\alpha\geq0$, the Tsallis $\alpha$-entropy of discrete
probability distribution is defined as
\begin{equation}
H_{\alpha}(P)=\frac{1}{1-\alpha}\left(\sum\nolimits_{j} p_{j}^{\alpha}-1\right)
 . \label{tsedf}
\end{equation}
In the limit $\alpha\searrow0$, we treat (\ref{tsedf}) as the number
of non-zero probabilities minus $1$. In information theory, the
above entropic form was discussed in the papers \cite{havrda,darozy}
beyond the scope of non-extensive statistical physics. It is useful
to put the $\alpha$-logarithm of variable $x>0$,
\begin{equation}
\ln_{\alpha}(x)=
\begin{cases}
 \frac{x^{1-\alpha}-1}{1-\alpha} \>, & \text{ for } 0<\alpha\neq1\, , \\
 \ln{x} \, , & \text{ for } \alpha=1\,.
\end{cases}
\label{lnal}
\end{equation}
This function allows us to rewrite (\ref{tsedf}) in two equivalent
ways, namely
\begin{equation}
H_{\alpha}(P)=-\sum\nolimits_{j}p_{j}^{\alpha}\,\ln_{\alpha}(p_{j})
=\sum\nolimits_{j}p_{j}\,\ln_{\alpha}\!\left(\frac{1}{p_{j}}\right)
 . \label{tsaent2}
\end{equation}
It is now clear that the case $\alpha=1$ gives the standard Shannon
entropy $H_{1}(P)=-\sum_{j}p_{j}\ln{p}_{j}$. It will also helpful to
rewrite (\ref{tsaent2}) in the form
\begin{equation}
H_{\alpha}(P)=-\sum\nolimits_{j}\eta_{\alpha}(p_{j})
\, . \label{hpet}
\end{equation}
The latter deals with the function
\begin{equation}
\eta_{\alpha}(x)=\frac{x^{\alpha}-x}{1-\alpha}
\ , \label{etal}
\end{equation}
where $x\in[0,1]$ and $\alpha\geq0$. We will also use the index of
coincidence defined as
\begin{equation}
I(P)=\sum_{j=1}^{n} p_{j}^{2}
\, . \label{indf}
\end{equation}
Then the Tsallis $2$-entropy reads as $H_{2}(P)=1-I(P)$. The index
of coincidence appears in various questions of information theory.
In cryptography, for example, it gives a measure of the relative
frequency of symbols in a ciphertext sample \cite{mvov97}.
Inequalities between $I(P)$ and $H_{1}(P)$ were considered in
\cite{harr2001}. The resulting range is similar to the range
obtained in terms of the maximal probability and $H_{1}(P)$
\cite{feder94}. In effect, the results of that paper can be treated
as limiting cases of findings of Harremo\"{e}s and Tops{\o}e
\cite{harr2001}.

In the following, the Tsallis entropy will mainly be dealt with.
Nonetheless, the main results of the current study also concern the
R\'{e}nyi $\alpha$-entropy expressed as \cite{renyi61}
\begin{equation}
R_{\alpha}(P)=\frac{1}{1-\alpha}\,\ln\!\left(\sum\nolimits_{j} p_{j}^{\alpha}\right)
 . \label{reedf}
\end{equation}
For $\alpha\searrow0$, it leads to the logarithm of the number of
non-zero probabilities. The limit $\alpha\to1$ again gives the
Shannon entropy. Using the obvious connection written as
\begin{equation}
R_{\alpha}(P)=\frac{1}{1-\alpha}\,\ln\bigl[1+(1-\alpha)H_{\alpha}(P)\bigr]
\ , \label{reedt}
\end{equation}
each of Tsallis-entropy inequalities has a R\'{e}nyi-entropy
counterpart. In contrast to (\ref{tsedf}), the right-hand side of
(\ref{reedf}) is certainly concave only for $\alpha\in(0,1)$.
Convexity properties of R\'{e}nyi's entropies with orders $\alpha>1$
depend on dimensionality of probabilistic vectors \cite{bengtsson}.
For instance, the binary R\'{e}nyi $\alpha$-entropy is strictly
concave with respect to probabilistic vector for $0<\alpha\leq2$
\cite{benra78}. Basic properties of the R\'{e}nyi and Tsallis
entropies together with some of their physical applications are
discussed in \cite{ja04,bengtsson}.

Measurements of special types are indispensable in quantum
information science. Mutually unbiased bases, mutually unbiased
measurements and equiangular tight frames are important examples.
Entropic uncertainty relations for such measurements are widely used
in building feasible schemes to detect non-classical correlations.
The inequalities (\ref{hat1}) and (\ref{rat1}) allow us to improve
existing inequalities in terms of the Tsallis and R\'{e}nyi
entropies. It is instructive to recall basic facts concerning the
measurements of interest.

Let $\clb=\bigl\{|b_{i}\rangle\bigr\}$ and
$\clb^{\prime}=\bigl\{|b_{j}^{\prime}\rangle\bigr\}$ be two
orthonormal bases in $d$-dimensional Hilbert space $\hh_{d}$. These
bases are called mutually unbiased if and only if for all $i$ and
$j$,
\begin{equation}
\bigl|\langle{b}_{i}|b_{j}^{\prime}\rangle\bigr|=\frac{1}{\sqrt{d}}
\ . \nonumber
\end{equation}
The set $\cmb=\bigl\{\clb^{(1)},\ldots,\clb^{(M)}\bigr\}$ is a set
of mutually unbiased bases (MUBs), when each two entries of $\cmb$
are mutually unbiased. The states within MUBs are indistinguishable
in the following sense. If the two observables have unbiased
eigenbases, then the measurement of one observable reveals no
information about possible outcomes of the measurement of other. In
two dimensions, three eigenbases of the Pauli matrices are mutually
unbiased. Basic properties of MUBs are reviewed in \cite{bz10}.

Symmetric informationally complete measurements form another
especially important class of quantum measurements \cite{rbksc04}.
There are indications that such measurements exist in all
dimensions. At the same time, there is no universal method to build
them for all $d$. Hence, we obtain a reason to consider equiangular
tight frames. Only complex frames will be considered in the
following. A set of $n\geq{d}$ unit vectors $|\phi_{j}\rangle$ is
called a tight frame, when \cite{abfg19}
\begin{equation}
\sum_{j=1}^{n}\bigl|\langle\phi_{j}|\psi\rangle\bigr|^{2}
=S
\label{abframe}
\end{equation}
for all unit $|\psi\rangle\in\hh_{d}$. It is easy to see that
$S=n/d$. The tight frame is called equiangular, when there exists
$c>0$ such that
\begin{equation}
\bigl|\langle\phi_{i}|\phi_{j}\rangle\bigr|^{2}=c
\qquad (i\neq{j})
\, . \label{md2c}
\end{equation}
It can be shown that $cd=(n-d)/(n-1)$ and $n\leq{d}^{2}$. For any
equiangular tight frame (ETF), the resolution $\clf=\{\mf_{j}\}$ of
the identity reads as
\begin{equation}
\sum_{j=1}^{n}\mf_{j}=\pen_{d}
\, , \qquad
\mf_{j}=\frac{d}{n}\>
|\phi_{j}\rangle\langle\phi_{j}|
\, . \label{res}
\end{equation}
When the measured state is described by density matrix $\bro$ with
$\tr(\bro)=1$, the probability of $j$-th outcome is equal to
\begin{equation}
p_{j}(\clf;\bro)=\frac{d}{n}\,\langle\phi_{j}|\bro|\phi_{j}\rangle
\, . \label{prej}
\end{equation}
If there is an ETF with $n$ elements in $d$ dimensions, then an ETF
with $n$ elements exists in $n-d$ dimensions \cite{sustik,mixon}.
Even though existing lists of ETFs are rare, there is a way to build
new ETFs from known ones. For various applications of ETFs, see
\cite{caskut} and references therein.

A ETF with $n=d^{2}$, when exists, leads to the symmetric
informationally complete POVM (SIC-POVM). This measurements consists
of $d^{2}$ rank-one elements of the form
\begin{equation}
\nm_{j}=\frac{1}{d}\>|\phi_{j}\rangle\langle\phi_{j}|
\, . \label{nphj}
\end{equation}
Substituting $n=d^{2}$ into $cd=(n-d)/(n-1)$ and (\ref{md2c}) shows
that the normalized vectors $|\phi_{j}\rangle$ obey the condition
\begin{equation}
\bigl|\langle\phi_{i}|\phi_{j}\rangle\bigr|^{2}=\frac{1}{d+1}
\qquad (i\neq{j})
\, . \label{phj}
\end{equation}
The set $\cln=\{\nm_{j}\}$ is a symmetric informationally complete
measurement introduced in \cite{rbksc04}. For explicit constructions
of SIC-POVMs, see \cite{fhs2017,sflammia} and references therein.

The authors of \cite{kalev14} proposed a concept of mutually
unbiased measurements (MUMs). Let $\cle=\{\me_{i}\}$ and
$\cle^{\prime}=\{\me_{j}^{\prime}\}$ be two POVM measurements, each
with $d$ elements. We assume that POVM elements satisfy
\begin{align}
& \tr(\me_{i})=\tr(\me_{j}^{\prime})=1
\ , \label{tmn1}\\
& \tr(\me_{i}\me_{j}^{\prime})=\frac{1}{d}
\ . \label{dmn1}
\end{align}
The following fact follows from the assumptions. The
Hilbert--Schmidt product of two elements from the same POVM can be
described in terms of a single parameter $\varkappa$ \cite{kalev14}:
\begin{equation}
\tr(\me_{i}\me_{j})=\delta_{ij}{\,}\varkappa
+(1-\delta_{ij}){\>}\frac{1-\varkappa}{d-1}
\ . \label{mjmk}
\end{equation}
General bounds on the parameter $\varkappa$ are written as
$1/d\leq\varkappa\leq1$ \cite{kalev14}. The set
$\mgb=\bigl\{\cle^{(1)},\ldots,\cle^{(M)}\bigr\}$ is a set of MUMs
of the efficiency $\varkappa$, when each two measurements obey the
above properties. It turns out that we can reach the aim to build a
complete set of $d+1$ mutually unbiased measurements in $d$
dimensions \cite{kalev14}.

In a similar vein, symmetric informationally complete measurements
with elements of arbitrary rank can be treated. A general SIC-POVM
$\cleg=\{\hlm_{j}\}$ is characterized by a single parameter
\cite{gour2014}, which we denote by $\theta$. It consists of $d^{2}$
positive operators $\hlm_{j}$ such that \cite{gour2014}
\begin{equation}
\tr\bigl(\hlm_{j}\hlm_{j}\bigr)=\theta
 \label{ficnd}
\end{equation}
for all $j=1,\ldots,d^{2}$ and
\begin{equation}
\tr\bigl(\hlm_{i}\hlm_{j}\bigr)=\frac{1-\theta{d}}{d(d^{2}-1)}
\qquad (i\neq{j})
\, . \label{secnd}
\end{equation}
That is, the pairwise Hilbert-Schmidt product is one and the same
for all pairs of POVM elements. It is easy to see that
$\tr(\hlm_{j})=d^{-1}$ for all $j=1,\ldots,d^{2}$. The determining
parameter is restricted as \cite{gour2014}
\begin{equation}
\frac{1}{d^{3}}<\theta\leq\frac{1}{d^{2}}
\ . \label{resu}
\end{equation}
The strict inequality on the left is necessary, since the choice
$\theta{d}^{3}=1$ leads to $\hlm_{j}=d^{-2}\pen_{d}$ for all
$j=1,\ldots,d^{2}$ \cite{gour2014}. This measurement formally
allowed is of no interest in practice.

\section{Statement of the problem}\label{sec3}

Let us formulate explicitly the problem addressed in this paper. The
so-called information diagrams are a powerful tool to examine
relations between the Shannon entropy and the index of coincidence
\cite{harr2001}. Here, we aim to apply information diagrams to
Tsallis $\alpha$-entropies. To each probability distribution, one
assigns the two values: the index of coincidence and the entropy of
interest. These values are respectively interpreted as the abscissa
and ordinate of the plane. Hence, each probability distribution is
pictured by a point in the plane. Note that such sets of points
essentially depend on the allowed maximal number of non-zero
probabilities. In ideal, we wish to describe exactly the boundaries
of the above sets of points. In general, the problem to determine
the best bounding curves is very difficult. Nevertheless, particular
results of significant interest could be derived. For example, the
following fact was observed for the Shannon entropy \cite{harr2001}.
In effect, the theoretically best lower-bounding curve is not
smooth. Even if this curve remains indescribable analytically, there
is a way to improve an obvious estimation of the Shannon entropy
from below. It is reached by replacing the corresponding smooth
curve by a suitable polygonal line \cite{harr2001}. Let us consider
the latter in more detail. It will be helpful to deal with the
$\alpha$-entropy (\ref{tsedf}) directly. The Shannon entropy is then
taken only as a special choice of the entropic parameter.

It immediately follows from Jensen's inequality that, for $\alpha\in[0,2]$,
\begin{equation}
H_{\alpha}(P)\geq\ln_{\alpha}\!\left(\frac{1}{I(P)}\right)
 . \label{hapi}
\end{equation}
This fact allows one to derive uncertainty relations when the actual
index of coincidence can be estimated \cite{rastmubs}. For
(\ref{ukdf}), it holds that that
\begin{equation}
I(U_{k})=\frac{1}{k}\equiv{x}_{k}
\ , \qquad
H_{\alpha}(U_{k})=\ln_{\alpha}(k)
\, . \label{infdkk}
\end{equation}
So, one divides the interval $x\in[0,1]$ by the values $x_{k}=k^{-1}$
with integer $k\geq2$. The points
$\bigl(x_{k},\ln_{\alpha}(k)\bigr)$ lie on the smooth line
$x\mapsto\ln_{\alpha}\bigl(x^{-1}\bigr)$ used in the right-hand side of
(\ref{hapi}). Can this inequality be made more tight? The answer is
certainly positive for the Shannon entropy. It was shown by
Harremo\"{e}s and Tops{\o}e \cite{harr2001} that the line
$x\mapsto-\ln{x}$ can be replaced with the polygonal line connecting
the points $(x_{k},\ln{k})$ with integer $k$. Suppose for a time
that this conclusion still holds for the Tsallis $\alpha$-entropy.
The straightforward line between the incident points of the form
$\bigl(x_{k},\ln_{\alpha}(k)\bigr)$ reads as
\begin{equation}
x\mapsto{k}(k+1)\bigl[\,\ln_{\alpha}(k)-\ln_{\alpha}(k+1)\bigr](x-x_{k})+\ln_{\alpha}(k)
=a_{\alpha{k}}-b_{\alpha{k}}x
\, , \label{segm}
\end{equation}
where the coefficients are expressed as
\begin{equation}
a_{\alpha{k}}=(k+1)\ln_{\alpha}(k+1)-k\ln_{\alpha}(k)
\, , \qquad
b_{\alpha{k}}=k(k+1)\bigl[\,\ln_{\alpha}(k+1)-\ln_{\alpha}(k)\bigr]
\, . \label{abak}
\end{equation}
The polygonal line of interest consists of segments of the form
(\ref{segm}). Extending the ideas of \cite{harr2001}, we further
take the function
\begin{equation}
f_{\alpha{k}}(x)=\eta_{\alpha}(x)-a_{\alpha{k}}x+b_{\alpha{k}}x^{2}
\, . \label{fap}
\end{equation}
Substituting probabilities, we immediately obtain
\begin{equation}
F_{\alpha{k}}(P)=\sum_{j=1}^{n}f_{\alpha{k}}(p_{j})=H_{\alpha}(P)-a_{\alpha{k}}+b_{\alpha{k}}\,I(P)
\, . \label{pfap}
\end{equation}
For $\alpha=1$, the quantity (\ref{pfap}) is shown to be
non-negative \cite{harr2001}. In this paper, we aim to prove the
same for $\alpha\in[0,2]$. Thereby, we obtain improved lower bounds
expressed in terms of the index of coincidence.

It is instructive to discuss the least cases $\alpha=0$ and
$\alpha=2$ right now. The latter is obvious due to
$a_{2k}=b_{2k}=1$, whence
\begin{equation}
F_{2k}(P)=H_{2}(P)-1+I(P)\equiv0
 \label{f2k0}
\end{equation}
by definition of the Tsallis $2$-entropy. For $\alpha=2$, the
quantity (\ref{pfap}) is zero identically. For $\alpha\searrow0$,
one has $a_{0k}=2k$ and $b_{0k}=k^{2}+k$. It will be shown in the
next section that the task reduces to prove
$F_{\alpha{k}}(P_{*})\geq0$, where $P_{*}$ is a mixture of $U_{k}$
and $U_{k+1}$. This mixture can be assumed to be non-trivial, since
for each of the uniform distributions $U_{k}$ and $U_{k+1}$ the
quantity (\ref{pfap}) vanishes by construction. With $k+1$ non-zero
probabilities, we have $H_{0}(P_{*})=k$ and
\begin{equation}
F_{0k}(P_{*})=(k^{2}+k)I(P_{*})-k\geq0
 \label{f0k0}
\end{equation}
due to $I(P_{*})\geq(k+1)^{-1}$. Thus, we have check
$F_{\alpha{k}}(P)\geq0$ in the least points of the parameter
interval $\alpha\in[0,2]$.

\begin{figure*}
\centering \includegraphics[width=8.2cm]{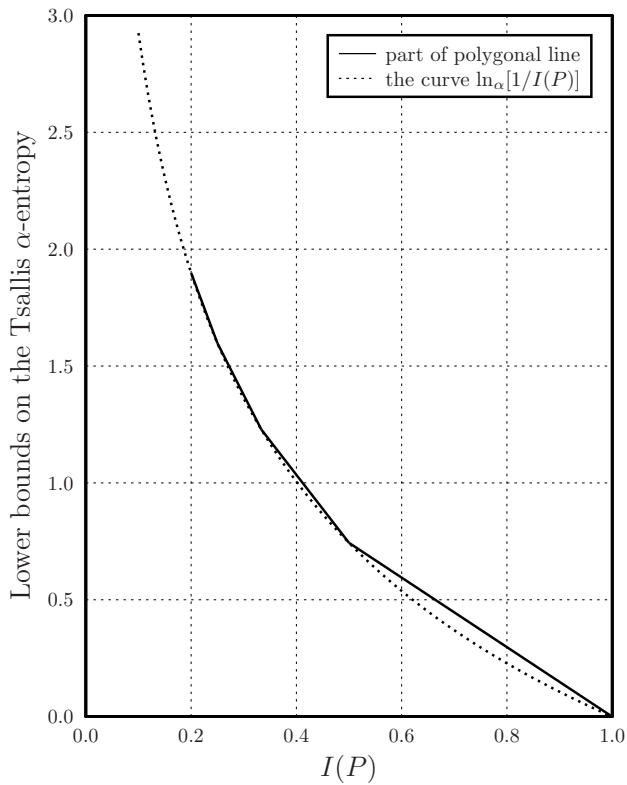}
\caption{\label{fig1} Two lower bounds on the Tsallis
$\alpha$-entropy for $\alpha=0.8$; polygonal line is restricted to
the case with no more than five non-zero probabilities.}
\end{figure*}

To illustrate the aim, both the smooth line and the polygonal one
are shown in Fig. \ref{fig1}. It shows how the polygonal line
improves an estimation from below that follows from (\ref{hapi}). It
is seen that an improvement is most valuable within the range
$1/2<I(P)<1$. Nevertheless, this advantage is of essential interest,
for example, in the following sense. In applications of uncertainty
relations, we often need restrictions that hold for all states. For
instance, such restrictions are used to derive separability
conditions and steering inequalities. Further, state-independent
formulation of entropic relations are typically reached with
substituting a pure state. For many measurements of special interest
the index of coincidence is maximal for certain pure states.
Examples of measurements with special properties will be considered
in Sec. \ref{sec5}. Even if the resulting improvement mainly plays
its role within a restricted range, it is essential for schemes to
detect non-classical correlations. It is also seen in Fig.
\ref{fig1} that the two lower bounds become coinciding for
sufficiently small values of the index of coincidence. Here,
separate segments of the polygonal line are shortened as well.

This section will be concluded with an example of information
diagram, bounding curves of which are expressed analytically. This
case deals with the range of the maximal probability versus the
index of coincidence. Here, the maximal number of non-zero
probabilities should be fixed. Let $x\mapsto\varLambda_{p}(x)$ be a
piecewise smooth function such that
\begin{equation}
\varLambda_{p}(x)=\frac{1}{k}\,\biggl(1+\sqrt{\frac{kx-1}{k-1}}\>\biggr)
\, , \qquad
x\in[x_{k},x_{k-1}]
\, . \label{gpdf}
\end{equation}
By construction, this continuous function obeys
$\varLambda_{p}(x_{k})=x_{k}$. The following statement takes place.

\newtheorem{tsap}{Theorem}
\begin{tsap}\label{thm1}
For each $P\in{M}_{+}^{1}(n)$ with $n\geq2$, it holds that
\begin{equation}
\varLambda_{p}\bigl(I(P)\bigr)\leq
\underset{1\leq{j}\leq{n}}{\max}\,p_{j}\leq
\frac{1}{n}\,\Bigl(1+\sqrt{n-1}\,\sqrt{nI(P)-1}\,\Bigr)
\, . \label{pat0}
\end{equation}
\end{tsap}

{\bf Proof.} We begin with the inequality on the right, though it
was actually proved in \cite{rastmubs}. The argument is presented
here for completeness and reformulation, since lemma 3 of
\cite{rastmubs} is posed in other terms. Let $p_{1}$ denote the
maximal probability. By convexity of the function
$\xi\mapsto\xi^{2}$, one has
\begin{equation}
I(P)\geq{p}_{1}^{2}+\frac{(1-p_{1})^{2}}{n-1}=\frac{np_{1}^{2}-2p_{1}+1}{n-1}
\ , \label{p1n}
\end{equation}
with equality if and only if $p_{j}=(n-1)^{-1}(1-p_{1})$ for all
$j=2,\ldots,n$. We further rewrite (\ref{p1n}) as
\begin{equation}
(np_{1}-1)^{2}\leq(n-1)\bigl[nI(P)-1\bigr]
\, \label{qnp}
\end{equation}
and the inequality on the right in (\ref{pat0}) follows due to
$np_{1}\geq1$. This inequality is saturated, if and only if all the
probabilities except possibly the maximal one are equal.

To prove the inequality on the left, more sophisticated reasons are
needed. It follows from the normalization that
\begin{equation}
I(P)\leq{p}_{1}\sum_{j=1}^{n}p_{j}=p_{1}
\, . \label{ip1}
\end{equation}
The latter proves the inequality on the left in (\ref{pat0}) for the
points $I(P)=x_{k}$. Suppose that $x_{k}<I(P_{0})<x_{k-1}$ for the
given distribution $P_{0}$, whence $x_{k}<p_{1}$. Let us seek for
the maximum of $I(P)$ at the given maximal probability $p_{1}$. The
task is equivalent to maximize the function
\begin{equation}
I(P)-p_{1}^{2}=\sum_{j=2}^{n}p_{j}^{2}
 \label{fp2n}
\end{equation}
under the restrictions $0\leq{p}_{j}\leq{p}_{1}$ for $j=2,\ldots,n$ and
\begin{equation}
\sum_{j=2}^{n}p_{j}=1-p_{1}
\, , \label{sp2n}
\end{equation}
Replacing the latter with $\sum_{j=2}^{n}p_{j}\leq1-p_{1}$, we get a
non-empty polyhedral convex set. It is well known that the maximum
of a convex function relative to such set is attained at one of the
extreme points of this set (see, e.g., section 32 of \cite{rockaf}).
We can actually restrict a consideration to those extreme points
that satisfy (\ref{sp2n}). Indeed, the function (\ref{fp2n})
increases with each non-zero probability at the fixed others. The
imposed restrictions are invariant under any permutation of
$p_{2},\ldots,p_{n}$. Up to a permutation, each of the points of
interest leads to probability distribution $P_{\max}$ containing
$k-1$ probabilities equal to $p_{1}$ and one probability
$1-(k-1)p_{1}<p_{1}$. Hence, the two indices of coincidence are such
that
\begin{align}
&I(P_{0})\leq{I}(P_{\max})=(k-1)p_{1}^{2}+\bigl[1-(k-1)p_{1}\bigr]^{2}
=k(k-1)p_{1}^{2}-2(k-1)p_{1}+1
\, , \nonumber\\
&kI(P_{0})-1\leq
{k}I(P_{\max})-1=(k-1)(kp_{1}-1)^{2}
\, . \label{fpkn}
\end{align}
Combining the latter with $1<kp_{1}$ results in
$\varLambda_{p}\bigl(I(P_{0})\bigr)\leq{p}_{1}$. With the probability
distribution $P_{\max}$, the inequality on the left in (\ref{pat0})
is saturated. $\blacksquare$

\begin{figure*}
\centering \includegraphics[width=8.2cm]{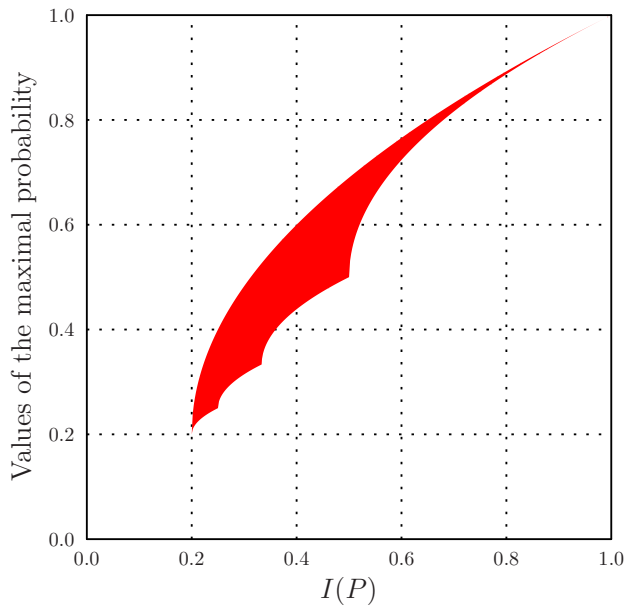}
\caption{\label{fig2} The allowed values of the
maximal probability versus the index of coincidence for $n=5$.}
\end{figure*}

The statement of Theorem \ref{thm1} described the exact range,
within which the maximal probability does vary at the given index of
coincidence. For $n=2$, there is one-to-one correspondence between
$I(P)$ and $\max{p}_{j}$, so that the diagram collapses into the
arc. Here, the two-sided estimate (\ref{pat0}) reduces to
\begin{equation}
\underset{1\leq{j}\leq{2}}{\max}\,p_{j}=
\frac{1+\sqrt{2I(P)-1}}{2}
\ . \nonumber
\end{equation}
Information diagrams for the Shannon entropy versus the maximal
probability were considered in \cite{feder94}. The authors of
\cite{harr2001} analyzed diagrams for the Shannon entropy versus the
index of coincidence. The two-sided estimate (\ref{pat0})
complements the mentioned results. Apparently, diagrams for
$\max{p}_{j}$ versus $I(P)$ are the simplest case to be resolved.
However, such diagrams do not seem to have been previously
recognized in the literature. The allowed values of the maximal
probability are shown in Fig. \ref{fig2} for the case
$P\in{M}_{+}^{1}(5)$. Here, the left- and right-hand sides of
(\ref{pat0}) are seen as the lower and upper-bounding curves of the
diagram. An enhancement due to the use of
$\varLambda_{p}\bigl(I(P)\bigr)$ instead of $I(P)$ becomes
negligible for sufficiently small values of the index of
coincidence. In contrast, the lower bound is improved essentially
for values between one half and one. Applications of the two-sided
estimate (\ref{pat0}) will be discussed in Sec. \ref{sec5}.

\section{Inequalities between entropies and the index of coincidence}\label{sec4}

To prove non-negativity of the quantity (\ref{pfap}), we recall the
lemma of replacement \cite{harr2001}. It holds for any
``concave/convex function with inflection point''. The latter refers
to a function $x\mapsto{f}(x)$ with $f(0)=0$ for which there exists
$0\leq\xi\leq1$ such that the restriction of $f$ to $[0,\xi]$ is
concave and the restriction of $f$ to $[\xi,1]$ is convex. For
$\alpha\in(0,2)$, the formula (\ref{fap}) gives a ``concave/convex
function with inflection point'' as defined in \cite{harr2001}.
Indeed, we have
\begin{equation}
f_{\alpha{k}}^{\prime\prime}(x)=\eta_{\alpha}^{\prime\prime}(x)+2b_{\alpha{k}}
=2b_{\alpha{k}}-\alpha{x}^{\alpha-2}
\, . \label{fap2}
\end{equation}
For $\alpha=1$, the second derivative reads as
\begin{equation}
f_{1k}^{\prime\prime}(x)=2b_{1k}-\frac{1}{x}
\ . \label{fap21}
\end{equation}
Namely this case was considered in \cite{harr2001}. The inflection
point is defined by vanishing (\ref{fap2}), so that
\begin{equation}
\xi_{\alpha{k}}=\left(\frac{2b_{\alpha{k}}}{\alpha}\right)^{\!1/(\alpha-2)}
 . \label{xiak}
\end{equation}
This value is strictly positive. It can be shown that \cite{harr2001}
\begin{equation}
\frac{1}{2(k+1)}<\xi_{1k}<\frac{1}{2k}
\ . \label{xi1k}
\end{equation}
For $0<\alpha<2$, the second derivative (\ref{fap2}) is strictly
negative for sufficiently small $x>0$. In a left vicinity of the
point $x=1$, this quantity becomes positive in view of
$2b_{\alpha{k}}>\alpha$. Indeed, we can write
\begin{equation}
\ln_{\alpha}(k+1)-\ln_{\alpha}(k)=
\int_{k}^{k+1}\frac{\xdif{x}}{x^{\alpha}}
>\int_{k}^{k+1}\frac{\xdif{x}}{x^{2}}=\frac{1}{k(k+1)}
\ , \nonumber
\end{equation}
whenever $\alpha\in(0,2)$ and $k\geq1$. For $\alpha\in(0,2)$, thereby,
the second derivative (\ref{fap21}) changes its sign inside the
interval $[0,1]$, which herewith contains the value
$\xi_{\alpha{k}}$. According to the first part of the lemma of
replacement \cite{harr2001}, there exist an integer $q\geq1$ and a
mixture $P_{*}$ of $U_{q}$ and $U_{q+1}$ such that
\begin{equation}
F_{\alpha{k}}(P)\geq{F}_{\alpha{k}}(P_{*})
\, . \label{pemq}
\end{equation}
An auxiliary inequality will also be required in the proof of the
main result. This inequality is posed as follows.

\newtheorem{galp}{Lemma}
\begin{galp}\label{lem1}
For real $\alpha\in(1,2)$ and integer $k\geq1$, it holds that
\begin{equation}
g_{k}(\alpha)=\frac{\alpha}{k+1}+2\biggl(1+\frac{1}{k}\biggr)^{\!1-\alpha}\!-2>0
\, . \label{glp1}
\end{equation}
\end{galp}

{\bf Proof.} We first note that $g_{k}(2)=0$ and
\begin{equation}
\frac{\xdif^{2}g_{k}(\alpha)}{\xdif\alpha^{2}}=2\biggl(1+\frac{1}{k}\biggr)^{\!1-\alpha}\!\left[\,\ln\biggl(1+\frac{1}{k}\biggr)\right]^{2}
>0
\end{equation}
for $k\geq1$. Thus, the function of interest is strictly convex. The
first derivative reads as
\begin{equation}
\frac{\xdif{g}_{k}(\alpha)}{\xdif\alpha}=\frac{1}{k+1}-2\biggl(1+\frac{1}{k}\biggr)^{\!1-\alpha}\!\ln\biggl(1+\frac{1}{k}\biggr)
\, , \nonumber
\end{equation}
whence
\begin{equation}
\biggl(1+\frac{1}{k}\biggr)\left.
\frac{\xdif{g}_{k}(\alpha)}{\xdif\alpha}
\,\right|_{\alpha=2}=\frac{1}{k}-2\ln\biggl(1+\frac{1}{k}\biggr)
<\frac{1-k}{k(k+1)}\leq0
\, . \nonumber
\end{equation}
The latter was used in \cite{harr2001} and follows from the
inequality
\begin{equation}
\ln\biggl(1+\frac{1}{k}\biggr)=\int_{0}^{1/k}\frac{\xdif\tau}{1+\tau}>
\int_{0}^{1/k}\frac{\xdif\tau}{(1+\tau)^{2}}=
\frac{1}{k+1}
\ . \nonumber
\end{equation}
It is strict with a strictly positive upper limit of integration.
So, one has a convex function intersecting the abscissa axis at the
point $\alpha=2$ with a strictly negative slope. Therefore, it is
strictly positive for all $\alpha\in(1,2)$. $\blacksquare$

We are now ready to establish the key result of this section. The
following statement takes place.

\newtheorem{tsaht}[tsap]{Theorem}
\begin{tsaht}\label{thm2}
Let $P\in{M}_{+}^{1}(\mnn)$, and let $x\mapsto{L}(x)$ be a piecewise
linear function such that
\begin{equation}
L_{\alpha}(x)=a_{\alpha{k}}-b_{\alpha{k}}x
\, , \qquad
x\in[x_{k+1},x_{k}]
\, , \label{eldf}
\end{equation}
where the coefficients are defined by (\ref{abak}). For real
$\alpha\in[0,2]$, it holds that
\begin{equation}
H_{\alpha}(P)\geq{L}_{\alpha}\bigl(I(P)\bigr)
\, . \label{hat01}
\end{equation}
\end{tsaht}

{\bf Proof.} It must be shown that $F_{\alpha{k}}(P)\geq0$ for
arbitrary probability distribution $P$. By the lemma of replacement,
we will prove the aim for $P_{*}$ being a mixture of $U_{q}$ and
$U_{q+1}$ with some $q\geq1$. For $q\neq{k}$, the desired inequality
follows from (\ref{hapi}). In fact, the line
$x\mapsto{a}_{\alpha{k}}-b_{\alpha{k}}x$ is built so that it goes
over $x\mapsto\ln_{\alpha}\bigl(x^{-1}\bigr)$ only between the
points $x_{k+1}$ and $x_{k}$. Hence, the question of interest
reduces to the case $q=k$.

Thus, we deal with the inequality
$F_{\alpha{k}}(P)\geq{F}_{\alpha{k}}(P_{*})$, where $P_{*}$ is a
mixture of $U_{k}$ and $U_{k+1}$. We further assume $0<\alpha<2$,
since the cases $\alpha=0$ and $\alpha=2$ were resolved as
(\ref{f0k0}) and (\ref{f2k0}), respectively. It is straightforward
to check that
\begin{align}
H_{\alpha}\bigl(xU_{k}+(1-x)U_{k+1}\bigr)&=
\frac{1}{1-\alpha}
\left\{k\biggl(\frac{k+x}{k(k+1)}\biggr)^{\!\alpha}+\biggl(\frac{1-x}{k+1}\biggr)^{\!\alpha}-1\right\}
 , \label{haxx}\\
I\bigl(xU_{k}+(1-x)U_{k+1}\bigr)&=
\frac{k+x^{2}}{k(k+1)}
\, . \label{iaxx}
\end{align}
It is the aim here to show that the following function cannot be
negative,
\begin{equation}
\Phi_{\alpha{k}}(x)=F_{\alpha{k}}(P_{*})=H_{\alpha}\bigl(xU_{k}+(1-x)U_{k+1}\bigr)-
a_{\alpha{k}}+b_{\alpha{k}}\,I\bigl(xU_{k}+(1-x)U_{k+1}\bigr)
\, . \label{phidf}
\end{equation}
By construction, one has $\Phi_{\alpha{k}}(0)=0$ and
$\Phi_{\alpha{k}}(1)=0$. The first derivative reads as
\begin{equation}
\Phi_{\alpha{k}}^{\prime}(x)=
\frac{\alpha}{1-\alpha}\>\frac{k^{1-\alpha}(k+x)^{\alpha-1}-(1-x)^{\alpha-1}}{(k+1)^{\alpha}}
+2x\bigl[\,\ln_{\alpha}(k+1)-\ln_{\alpha}(k)\bigr]
\, . \label{phidf1}
\end{equation}
We also obtain $\Phi_{\alpha{k}}^{\prime}(0)=0$ for $0<\alpha<2$ and
$\Phi_{\alpha{k}}^{\prime}(1)=-\infty$ for $0<\alpha<1$. It is
already known that $\Phi_{1k}^{\prime}(1)=-\infty$ \cite{harr2001}.
Due to the inequality (\ref{glp1}), for $1<\alpha<2$ one has
\begin{equation}
(1-\alpha)k^{\alpha-1}\Phi_{\alpha{k}}^{\prime}(1)=
\frac{\alpha}{k+1}+2\biggl(1+\frac{1}{k}\biggr)^{\!1-\alpha}\!-2>0
\nonumber
\end{equation}
and, herewith, $\Phi_{\alpha{k}}^{\prime}(1)<0$. The second
derivative then reads as
\begin{equation}
\Phi_{\alpha{k}}^{\prime\prime}(x)=2\bigl[\,\ln_{\alpha}(k+1)-\ln_{\alpha}(k)\bigr]
-\frac{\alpha}{(k+1)^{\alpha}}\>\bigl[k^{1-\alpha}(k+x)^{\alpha-2}+(1-x)^{\alpha-2}\bigr]
\, , \label{phippx}
\end{equation}
whence
\begin{equation}
\Phi_{\alpha{k}}^{\prime\prime}(0)=\frac{2}{1-\alpha}
\>\Bigl[(k+1)^{1-\alpha}-k^{1-\alpha}\Bigr]-\frac{\alpha(k+1)^{1-\alpha}}{k}
=(k+1)^{1-\alpha}
\left\{\frac{2(1+\tau)^{\alpha-1}-2}{\alpha-1}-\alpha\tau\right\}
 , \label{phipp0}
\end{equation}
where $\tau=k^{-1}$. Using the Taylor formula with remainder in
Lagrange's form leads to
\begin{equation}
2\>\frac{(1+\tau)^{\alpha-1}-1}{\alpha-1}=
2\tau+(\alpha-2)\tau^{2}+\frac{1}{3}\,(\alpha-2)(\alpha-3)(1+c)^{\alpha-4}\tau^{3}
>2\tau+(\alpha-2)\tau^{2}
\end{equation}
with some $0<c<\tau$ and $\alpha\in(0,2)$. Hence, it holds that
\begin{equation}
(k+1)^{\alpha-1}\Phi_{\alpha{k}}^{\prime\prime}(0)>
2\tau+(\alpha-2)\tau^{2}-\alpha\tau
=(2-\alpha)\tau(1-\tau)\geq0
\, , \label{phipp00}
\end{equation}
provided that $0<\tau\leq1$ and $\alpha\in(0,2)$. Thus, we have
$\Phi_{\alpha{k}}^{\prime\prime}(0)>0$ for all $\alpha\in(0,2)$ and
$k\geq1$.

We have observed that $\Phi_{\alpha{k}}(0)=\Phi_{\alpha{k}}(1)=0$,
$\Phi_{\alpha{k}}^{\prime}(0)=0$,
$\Phi_{\alpha{k}}^{\prime\prime}(0)>0$ and
$\Phi_{\alpha{k}}^{\prime}(1)<0$. It now follows that
\begin{equation}
\Phi_{\alpha{k}}(x)\geq0
 \label{phig0}
\end{equation}
for all $\alpha\in(0,2)$ and $x\in[0,1]$. Assuming a strictly
negative value, the function of interest would have at least three
inflection points that is impossible. Indeed, each inflection point
can be reinterpreted as an intersection of the straight line
$X+Y=k+1$ with the curve
\begin{equation}
\biggl(\frac{A_{\alpha{k}}}{X}\biggr)^{\!2-\alpha}+\biggl(\frac{B_{\alpha{k}}}{Y}\biggr)^{\!2-\alpha}=1
\, , \qquad
A_{\alpha{k}}^{\alpha-2}=k^{\alpha-1}B_{\alpha{k}}^{\alpha-2}=\frac{2}{\alpha}\,(k+1)^{\alpha}\ln_{\alpha}\biggl(1+\frac{1}{k}\biggr)
\, . \label{curAB}
\end{equation}
Here, we substitute the abscissa $X=k+x$ into the equation
$\Phi_{\alpha{k}}^{\prime\prime}(x)=0$. For $\alpha\in(0,2)$, the
curve (\ref{curAB}) represents the graph of a convex function that
tends to come asymptotically along the lines $X=A_{\alpha{k}}$ and
$Y=B_{\alpha{k}}$. More than two intersections with the straight
line $X+Y=k+1$ are not possible in the first quadrant.
$\blacksquare$

The statement of Theorem \ref{thm2} gives an improvement of the
lower bound based on (\ref{hapi}). The obtained improvement is most
essential within the interval $x_{2}<I(P)<x_{1}$. This fact was
already mentioned with respect to the lines shown in Fig.
\ref{fig1}. It is often useful to mention the maximal number of
probabilities explicitly. For $P\in{M}_{+}^{1}(n)$ with $n\geq2$ and
$\alpha\in[0,2]$, one can rewrite (\ref{hat01}) in the form
\begin{equation}
H_{\alpha}(P)\geq
\underset{1\leq{k}\leq{n}-1}{\max}
\bigl\{a_{\alpha{k}}-b_{\alpha{k}}\,I(P)\bigr\}
\, . \label{hat1}
\end{equation}
For $\alpha=1$, the inequality (\ref{hat1}) reads as
\begin{equation}
H_{1}(P)\geq
\underset{1\leq{k}\leq{n}-1}{\max}
\bigl\{a_{1k}-b_{1k}\,I(P)\bigr\}
\, , \label{hat11}
\end{equation}
where $a_{1k}=(k+1)\ln(k+1)-k\ln{k}$,
$b_{1k}=k(k+1)\ln\bigl(1+k^{-1}\bigr)$. This inequality is one of
the main results of Harremo\"{e}s and Tops{\o}e \cite{harr2001}. It
was proved therein by two methods one of which is based on the lemma
of replacement and the other uses topological considerations. Using
this lemma, the statement of Theorem \ref{thm2} extends
(\ref{hat11}) to Tsallis $\alpha$-entropies for all
$\alpha\in[0,2]$. Thereby, one obtains an inequality for R\'{e}nyi
$\alpha$-entropies of order $\alpha\in[0,2]$, viz.
\begin{equation}
R_{\alpha}(P)\geq
\frac{1}{1-\alpha}\,\ln\Bigl[\,1+(1-\alpha)\underset{1\leq{k}\leq{n}-1}{\max}\bigl\{a_{\alpha{k}}-b_{\alpha{k}}\,I(P)\bigr\}\Bigr]
\, , \label{rat1}
\end{equation}
where $n$ denotes the number of non-zero probabilities. In the limit
$\alpha\searrow0$, the formula (\ref{rat1}) implies
$I(P)\geq{n}^{-1}$ due to $R_{0}(P)=\ln{n}$. The next section deals
with applications of the above inequalities to entropic uncertainty
relations. Nevertheless, the new inequalities (\ref{hat1}) and
(\ref{rat1}) can found use in classical information theory similarly
to the result (\ref{hat11}). This question could be a subject of
separate investigation.

\section{Uncertainty relations in terms of generalized entropies}\label{sec5}

In this section, the obtained inequalities will be applied to
formulate entropic uncertainty relations for quantum measurements of
several types. Entropic uncertainty relations are interesting in
their own rights as well as for applications in quantum information
science \cite{cbtw17}. Here, the number of probabilities follows
from the context. To each mutually unbiased basis
$\clb=\bigl\{|b_{j}\rangle\bigr\}$, we assign the corresponding
index of coincidence
\begin{equation}
I(\clb;\bro)=\sum_{j=1}^{d}p_{j}(\clb;\bro)^{2}
\, , \label{inbf}
\end{equation}
where $p_{j}(\clb;\bro)=\langle{b}_{j}|\bro|b_{j}\rangle$. The
following inequality was proved in \cite{molm09}. For the set $\cmb$
of $M$ mutually unbiased bases, the indices of coincidence obey
\begin{equation}
\sum_{\clb\in{\!\;}\cmb}I(\clb;\bro)\leq\tr(\bro^{2})+\frac{M-1}{d}
\ . \label{lwbm}
\end{equation}
It follows from (\ref{hat01}) that, for all $\alpha\in[0,2]$ and
$1\leq{k}\leq{d}-1$,
\begin{equation}
\frac{1}{M}\sum_{\clb\in{\!\;}\cmb}H_{\alpha}(\clb;\bro)\geq
\frac{1}{M}\sum_{\clb\in{\!\;}\cmb}L_{\alpha}\bigl(I(\clb;\bro)\bigr)
\geq{L}_{\alpha}\!\left(\,\sum_{\clb\in{\!\;}\cmb}\frac{I(\clb;\bro)}{M}\right)
 . \label{basm0}
\end{equation}
We used here that the function $x\mapsto{L}_{\alpha}(x)$ is convex.
Since it also decreases, combining (\ref{lwbm}) with (\ref{basm0})
leads to the uncertainty relation
\begin{equation}
\frac{1}{M}\sum_{\clb\in{\!\;}\cmb}H_{\alpha}(\clb;\bro)\geq
\underset{1\leq{k}\leq{d}-1}{\max}
\left\{
a_{\alpha{k}}-b_{\alpha{k}}\,\frac{\tr(\bro^{2}){\!\;}d+M-1}{Md}\,
\right\}
 . \label{basm1}
\end{equation}
The latter is an uncertainty relation in terms of the Tsallis
$\alpha$-entropy averaged over the set of $M$ MUBs. The case
$\alpha=1$ reduces (\ref{basm1}) to the uncertainty relation in
terms of the Shannon entropy. It is actually equivalent to the
inequality stated as theorem 2 of the paper \cite{molm09}. It is
natural to extend the above uncertainty relation to mutually
unbiased measurements. The following statement takes place.

\newtheorem{tmum}[tsap]{Theorem}
\begin{tmum}\label{thm3}
Let $\mgb=\bigl\{\cle^{(1)},\ldots,\cle^{(M)}\bigr\}$ be a set of
$M$ mutually unbiased measurements of the efficiency $\varkappa$ in
$d$ dimensions. For $\alpha\in[0,2]$ and arbitrary density matrix
$\bro$, it holds that
\begin{equation}
\frac{1}{M}\sum_{\cle\in\,\mgb}H_{\alpha}(\cle;\bro)\geq
\underset{1\leq{k}\leq{d}-1}{\max}
\left\{
a_{\alpha{k}}-b_{\alpha{k}}\,\frac{M-1}{Md}
-b_{\alpha{k}}\,\frac{1-\varkappa+(\varkappa{d}-1){\,}\tr(\bro^{2})}{M(d-1)}
\,\right\}
 , \label{bmu1}
\end{equation}
where the coefficients are defined by (\ref{abak}). For
$\alpha\in[1,2]$, it also holds that
\begin{equation}
\frac{1}{M}\sum_{\cle\in\,\mgb}R_{\alpha}(\cle;\bro)\geq
\frac{1}{1-\alpha}\,\ln\biggl[\,1+(1-\alpha)\underset{1\leq{k}\leq{d}-1}{\max}\biggl\{
a_{\alpha{k}}-b_{\alpha{k}}\,\frac{M-1}{Md}
-b_{\alpha{k}}\,\frac{1-\varkappa+(\varkappa{d}-1){\,}\tr(\bro^{2})}{M(d-1)}
\,\biggr\}\biggr]
 . \label{brmu1}
\end{equation}
\end{tmum}

{\bf Proof.} It was proved in \cite{rastosid} that
\begin{equation}
\sum_{\cle\in\,\mgb}I(\cle;\bro)\leq\frac{M-1}{d}+\frac{1-\varkappa+(\varkappa{d}-1){\,}\tr(\bro^{2})}{d-1}
\ . \label{ubp1}
\end{equation}
Similarly to (\ref{basm1}), the result (\ref{bmu1}) then follows from
(\ref{hat01}) and (\ref{ubp1}).

Note that the function
$x\mapsto(1-\alpha)^{-1}\ln\bigl[1+(1-\alpha)x]$ is increasing.
Combining this with (\ref{reedt}) allows one to write
\begin{equation}
\frac{1}{M}\sum_{\cle\in\,\mgb}R_{\alpha}(\cle;\bro)\geq
\frac{1}{M}\sum_{\cle\in\,\mgb}
\frac{1}{1-\alpha}\,\ln\bigl[\,1+(1-\alpha)L_{\alpha}\bigl(I(\cle;\bro)\bigr)\bigr]
\, . \label{brmu22}
\end{equation}
For $1\leq\alpha\leq2$, the function
$x\mapsto(1-\alpha)^{-1}\ln\bigl[1+(1-\alpha)x]$ is convex and
increasing, whence the composition
$x\mapsto(1-\alpha)^{-1}\ln\bigl[1+(1-\alpha)L_{\alpha}(x)\bigr]$ is
convex. To get (\ref{brmu1}), we combine this convexity with
(\ref{brmu22}) and decreasing of $x\mapsto{L}_{\alpha}(x)$.
$\blacksquare$

The inequalities (\ref{bmu1}) and (\ref{brmu1}) respectively give
new uncertainty relations for the averaged Tsallis and R\'{e}nyi
entropies. In dimension two, we have three mutually unbiased bases
--- the eigenbases of the Pauli matrices. For $\alpha\in[1,2]$ and
$M=1,2,3$, it holds that
\begin{equation}
\frac{1}{M}\sum_{m=1}^{M}R_{\alpha}\bigl(\clb^{(m)};\bro\bigr)\geq
\frac{1}{1-\alpha}\,\ln\biggl[\,1+(1-\alpha)\underset{1\leq{k}\leq{d}-1}{\max}\biggl\{
a_{\alpha{k}}-b_{\alpha{k}}\,
\frac{2\,\tr(\bro^{2})+M-1}{2M}
\,\biggr\}\biggr]
 . \label{brmu12}
\end{equation}
Since the binary R\'{e}nyi entropy is concave up to $\alpha=2$
\cite{benra78}, the inequality (\ref{brmu12}) deals with concave
entropies. Applying uncertainty relations to detect quantum
correlations, we need inequalities that hold for all states. The
right-hand side of each of the two inequalities (\ref{bmu1}) and
(\ref{brmu1}) decreases with growth of $\tr(\bro^{2})$. Hence, we
obtain state independent formulations
\begin{align}
\frac{1}{M}\sum_{\cle\in\,\mgb}H_{\alpha}(\cle;\bro)&\geq
\underset{1\leq{k}\leq{d}-1}{\max}
\left\{
a_{\alpha{k}}-b_{\alpha{k}}\,\frac{M+\varkappa{d}-1}{Md}
\,\right\}
 , \label{bmu1p}\\
\frac{1}{M}\sum_{\cle\in\,\mgb}R_{\alpha}(\cle;\bro)
&\geq\frac{1}{1-\alpha}\,\ln\biggl[\,1+(1-\alpha)\underset{1\leq{k}\leq{d}-1}{\max}\biggl\{
a_{\alpha{k}}-b_{\alpha{k}}\,\frac{M+\varkappa{d}-1}{Md}
\,\biggr\}\biggr]
 ,  \label{brmu1p}
\end{align}
the first for $\alpha\in[0,2]$ and the second for $\alpha\in[1,2]$.
Actually, the right-hand sides of (\ref{bmu1p}) and (\ref{brmu1p})
are obtained for a pure state by substituting $\tr(\bro^{2})=1$.

It was already mentioned that equiangular tight frames are easier to
build than SIC-POVMs. Related measurements are useful in detection
of entanglement and steerability \cite{rastetf}. The Tsallis-entropy
uncertainty relation for an ETF-based measurement is formulated as
follows.

\newtheorem{untr2}[tsap]{Theorem}
\begin{untr2}\label{thm4}
Let $n$ unit kets $|\phi_{j}\rangle$ form an ETF in $\hh_{d}$, and
let POVM $\clf$ be assigned to this frame by (\ref{res}). For
$\alpha\in[0,2]$ and arbitrary density matrix $\bro$, it holds that
\begin{align}
H_{\alpha}(\clf;\bro)&\geq
\underset{1\leq{k}\leq{n}-1}{\max}
\left\{
a_{\alpha{k}}-b_{\alpha{k}}\,\frac{Sc+(1-c)\,\tr(\bro^{2})}{S^{2}}
\,\right\}
\, . \label{est2}\\
R_{\alpha}(\clf;\bro)&\geq
\frac{1}{1-\alpha}\,\ln\biggl[\,1+(1-\alpha)\underset{1\leq{k}\leq{n}-1}{\max}\biggl\{a_{\alpha{k}}-b_{\alpha{k}}\,
\frac{Sc+(1-c)\,\tr(\bro^{2})}{S^{2}}
\,\biggr\}\biggr]
\, . \label{ert2}
\end{align}
\end{untr2}

{\bf Proof.} For the POVM assigned to an ETF, the index of
coincidence satisfies \cite{rastetf}
\begin{equation}
\frac{d^{2}}{n^{2}}\,
\sum_{j=1}^{n}\langle\phi_{j}|\bro|\phi_{j}\rangle^{2}
\leq
\frac{Sc+(1-c)\,\tr(\bro^{2})}{S^{2}}
\ . \label{inbr}
\end{equation}
Combining (\ref{hat1}) with (\ref{inbr}) completes the proof of
(\ref{est2}). By a parallel argument, the result (\ref{ert2})
follows from (\ref{rat1}) and (\ref{inbr}).
$\blacksquare$

Thus, we obtain new uncertainty relations in terms of the Tsallis
and R\'{e}nyi entropies for an ETF-based quantum measurement.
Similarly to the case of mutually unbiased measurements, the
state-independent formulations are obtained with $\tr(\bro^{2})=1$.
Then the inequalities (\ref{est2}) and (\ref{ert2}) read as
\begin{align}
H_{\alpha}(\clf;\bro)&\geq
\underset{1\leq{k}\leq{n}-1}{\max}
\biggl\{
a_{\alpha{k}}-b_{\alpha{k}}\,\frac{d^{2}-2d+n}{n^{2}-n}
\,\biggr\}
\, . \label{est2p}\\
R_{\alpha}(\clf;\bro)&\geq
\frac{1}{1-\alpha}\,\ln\biggl[\,1+(1-\alpha)\underset{1\leq{k}\leq{n}-1}{\max}\biggl\{a_{\alpha{k}}-b_{\alpha{k}}\,
\frac{d^{2}-2d+n}{n^{2}-n}
\,\biggr\}\biggr]
\, . \label{ert2p}
\end{align}
Here, we substituted $S=n/d$ and $cd=(n-d)/(n-1)$. For a SIC-POVM
with $n=d^{2}$ operators of the form (\ref{nphj}), the inequality
(\ref{inbr}) is replaced with \cite{rastetf}
\begin{equation}
I(\cln;\bro)=\frac{1+\tr(\bro^{2})}{d(d+1)}
\ . \label{inbsic}
\end{equation}
In the paper \cite{rastmubs}, this index of coincidence was
calculated and applied to derive uncertainty relations in terms of
generalized entropies. Due to (\ref{est2}) and (\ref{ert2}), the
following inequalities hold,
\begin{align}
H_{\alpha}(\cln;\bro)&\geq
\underset{1\leq{k}\leq{d}^{2}-1}{\max}
\biggl\{
a_{\alpha{k}}-b_{\alpha{k}}\,\frac{1+\tr(\bro^{2})}{d(d+1)}
\,\biggr\}
\, , \label{sest2}\\
R_{\alpha}(\cln;\bro)&\geq
\frac{1}{1-\alpha}\,\ln\biggl[\,1+(1-\alpha)\underset{1\leq{k}\leq{d}^{2}-1}{\max}\biggl\{a_{\alpha{k}}-b_{\alpha{k}}\,
\frac{1+\tr(\bro^{2})}{d(d+1)}
\,\biggr\}\biggr]
\, , \label{sert2}
\end{align}
where $\alpha\in[0,2]$. These inequalities improve the results given in \cite{rastmubs}. It
is natural to extend (\ref{sest2}) and (\ref{sert2}) to the case of a general SIC-POVM.

\newtheorem{gntr2}[tsap]{Theorem}
\begin{gntr2}\label{thm5}
Let general SIC-POVM $\cleg$ be characterized by the parameter
$\theta$. For $\alpha\in[0,2]$ and arbitrary density matrix $\bro$,
it holds that
\begin{align}
H_{\alpha}(\cleg;\bro)&\geq
\underset{1\leq{k}\leq{d}^{2}-1}{\max}
\left\{
a_{\alpha{k}}-b_{\alpha{k}}\,\frac{d(1-\theta{d})+(\theta{d}^{3}-1)\,\tr(\bro^{2})}{d(d^{2}-1)}
\,\right\}
 , \label{gst2}\\
R_{\alpha}(\cleg;\bro)&\geq
\frac{1}{1-\alpha}\,\ln\biggl[\,1+(1-\alpha)\underset{1\leq{k}\leq{d}^{2}-1}{\max}\biggl\{a_{\alpha{k}}-b_{\alpha{k}}\,
\frac{d(1-\theta{d})+(\theta{d}^{3}-1)\,\tr(\bro^{2})}{d(d^{2}-1)}
\,\biggr\}\biggr]
 , \label{rst2}
\end{align}
\end{gntr2}

{\bf Proof.} For a general SIC-POVM, one has \cite{rastsic}
\begin{equation}
I(\cleg;\bro)=\frac{d(1-\theta{d})+(\theta{d}^{3}-1)\,\tr(\bro^{2})}{d(d^{2}-1)}
\ . \label{icpr}
\end{equation}
Combining (\ref{hat1}) with (\ref{icpr}) immediately leads to
(\ref{gst2}). In a similar manner, the result (\ref{rst2}) follows
from (\ref{rat1}) and (\ref{icpr}).
$\blacksquare$

Thus, we derive new uncertainty relations in terms of the Tsallis
and R\'{e}nyi entropies for a general SIC-POVM. Similarly to the
above cases, state-independent formulations are obtained with
$\tr(\bro^{2})=1$, namely
\begin{align}
H_{\alpha}(\cleg;\bro)&\geq
\underset{1\leq{k}\leq{d}^{2}-1}{\max}
\left\{
a_{\alpha{k}}-b_{\alpha{k}}\,\frac{\theta{d}^{2}+1}{d(d+1)}
\,\right\}
 , \label{gst2p}\\
R_{\alpha}(\cleg;\bro)&\geq
\frac{1}{1-\alpha}\,\ln\biggl[\,1+(1-\alpha)\underset{1\leq{k}\leq{d}^{2}-1}{\max}\biggl\{a_{\alpha{k}}-b_{\alpha{k}}\,
\frac{\theta{d}^{2}+1}{d(d+1)}
\,\biggr\}\biggr]
 . \label{rst2p}
\end{align}

For a SIC-POVM or a general SIC-POVM, the index of coincidence is
known exactly due to (\ref{icpr}). Combining this with (\ref{pat0})
allows us to give a two-sided estimate on the min-entropy, viz.
\begin{equation}
2\ln{d}-\ln\Bigl(
1+\sqrt{\theta{d}^{3}-1}\,\sqrt{d\,\tr(\bro^{2})-1}
\,\Bigr)
\leq{R}_{\infty}(\cleg;\bro)
\leq{}\!-\ln\biggl[\,\varLambda_{p}\biggl(\frac{d(1-\theta{d})+(\theta{d}^{3}-1)\,\tr(\bro^{2})}{d(d^{2}-1)}\biggr)\biggr]
\, , \label{rnim}
\end{equation}
where the function $x\mapsto\varLambda_{p}(x)$ is defined by
(\ref{gpdf}). For the maximally mixed state
$\bro_{*}=d^{-1}\pen_{d}$, this two-sided estimate leads to
$R_{\infty}(\cleg;\bro_{*})=2\ln{d}$ as expected.

This section presented new uncertainty relations in terms of
generalized entropies for several quantum measurements. In
particular, new state-independent formulations were obtained. It was
already mentioned that such restrictions can be used to test
non-classical correlations. Applying the above results to derive
separability conditions and steering inequalities is rather the
subject of a separate investigation.

\section{Conclusions}\label{sec6}

Using information diagrams has led to new lower bounds on the
R\'{e}nyi and Tsallis entropies in terms of the index of
coincidence. The presented derivation used the lemma of replacement
given in \cite{harr2001} and applied therein to the Shannon entropy.
Thus, we have extended some results of that paper to generalized
entropies. In addition, the tight lower and upper bounds on the
maximal probability in terms of the index of coincidence were
formulated. New relations between mentioned informational
characteristics can be used in various questions. To illustrate
possible applications, new entropic uncertainty relations were
formulated for some measurements with interesting properties. In
particular, we considered mutually unbiased bases, symmetric
informationally complete measurements and measurement assigned to
equiangular tight frames. Further applications of the presented
inequalities in quantum information science are hoped to be
addressed in a future work.

\end{document}